\documentclass[10pt]{article}
\usepackage{amssymb}
\usepackage{graphicx}
\graphicspath{{figure/}} %find in the folder figure the figures

\usepackage[utf8]{inputenc}
\usepackage[francais, english]{babel} 
\usepackage[T1]{fontenc}
\usepackage{lmodern}
\usepackage{amsmath}
\usepackage{caption}

\usepackage{plain}
\usepackage{subcaption}
\usepackage{multirow}
\usepackage{gensymb}

\usepackage{multicol} % Used for the two-column layout of the document         
\usepackage{tabularx}
\usepackage{cite}
\usepackage[compact]{titlesec}
\titlespacing*{\section}{0pt}{0pt plus 3pt}{3pt}
\titlespacing*{\subsection}
{0pt}{5.5ex plus 1ex minus .2ex}{4.3ex plus .2ex}

\usepackage[a4paper]{geometry}%  page type
\usepackage[cm]{fullpage} % use full page
\usepackage{float} % Required for tables and figures in the multi-column environment - they need to be placed in specific locations with the [H] (e.g. \begin{table}[H])
\usepackage[labelfont=it]{caption}
\usepackage{array}
\usepackage{authblk} %pour les affiliations

\usepackage{xcolor,soul}
\usepackage{mathpazo} % Palatino font

\begin{document}

%----------------------------------------------------------------------------------------
%	TITLE PAGE
%----------------------------------------------------------------------------------------

%\setkeys{Gin}{draft}
%============ titre ==========================
\title{\fontsize{20pt}{9pt}\selectfont\textbf{ Technical Note: A Novel Electromagnetic-Tracked Scintillation Dosimeter for Accurate \textit{In Vivo} Dosimetry in HDR Brachytherapy.}}
\author[1,2]{{Daline Tho}}
\author[1,2]{Luc Beaulieu} 
\affil[1]{ Département de radio-oncologie et Centre de recherche du CHU de Québec, CHU de Québec,Canada}
\affil[2]{ Département de physique, de génie physique et d'optique, et Centre de recherche sur le cancer, Université Laval, Canada}
\sloppy

%===========================================

\maketitle
Corresponding author: Luc Beaulieu\\
email: Luc.Beaulieu@phy.ulaval.ca 
%===============resume=======================

\begin{abstract}

 \textbf{Purpose:} Brachytherapy is a treatment modality which delivers large doses of radiation in a reduced number of visits. Since large dose are administrated to the patient, ensuring the right dose is delivered is highly critical. This study presents a first step in solving the standing issue of accurately knowing the dosimeter position at all time during \textit{in vivo} dosimetry. In this work, an energy independent dosimeter, namely plastic scintillation detector, is coupled to an electromagnetic (EM) sensor having sub-mm positional accuracy for real-time tracking of the dosimeter position. However, adding an EM sensor adds materials in the path to the scintillator and thus could potentially perturb the dose measurements. \textbf{Methods:} To confirm the perturbation presence, 4 different sensors were placed in front of the scintillator so the radiation does not arrive to it directly. Variation of the distance between the sensor and the scintillator was used to quantify the effect on the signal at 0$\degree$ and 90$\degree$. To test the signal's angular dependence for each sensor, the signal  measurement were taken from 0$\degree$ to 90$\degree$ with 10$\degree$ increment.  \textbf{Results:}The 5PCBDOF sensor showed an increased signal of almost 20 $\%$ with increasing beam angle. Sensor 5DOF, 5DOFthin and 6DOF showed no significant angle dependance. The 6DOF and 5DOFthin sensor's cable revealed no extra signal attenuation. The latter gives a smaller overall attenuation. Therefore, the 5DOFthin is chosen to be part of the novel dosimeter construction. It has a jitter error of $\pm$ 0.06 mm and a reproductibility of $\pm$  0.008 mm. In the optimal operating range, the average positional uncertainty is less than 0.2 mm. Average angle errors are at most of 1.1$\degree$. \textbf{Conclusion:} It is feasible to integrate an EM tracking sensor to an energy independent plastic scintillation dosimeter with minimal impact to the collected signal as well as sufficient positional accuracy to keep dose uncertainty below 5$\%$.

 \end{abstract}
%================INTRODUCTION===========================

\section {Introduction}

Brachytherapy is a treatment modality which delivers large doses of radiation in a reduced number of visits \cite{Devlin:2007aa}. Because large doses are administrated to the patient, ensuring that the right dose is delivered is highly critical. As such  \textit{ in vivo} dose measurements is a key QC/QA activity \cite{Tanderup:2013aa}. Numerous works have tackled this issue using different detectors as underlined in a recent Vision 20/20 manuscript by Tanderup et al\cite{Tanderup:2013aa}. Two major problems with those detectors are related to the use of non water-equivalent material and the incapacity of making real-time measurement \cite{Andersen:2009aa,Bloemen-van-Gurp:2009aa, Ciesielski:2003aa}. The former concern is usually responsible for detector energy dependance and the latter points out the fact that some treatment errors can be invisible to integral dose measurements \cite{Tanderup:2013aa}. To address the brachytherapy needs for a complete real-time \textit{in vivo} dosimetry, a combination of an electromagnetic tracking (EMT) technology  with a plastic scintillation detector (PSD) is proposed here.

EMT systems are widely used in the medical field such as image-guided surgery \cite{Baszynski2010, Franz:2014aa}. Specifically, the Aurora EMT system uses a field generator which produces an electromagnetic field gradient within a cubic volume of 50 cm side, inside of which, a small sensor can be placed. The system uses mutual induction to compute the sensor's position \cite{Seiler:2000aa}. Its use has been proposed for real-time guidance of needles and catheters in high-dose rate Ir-192 brachytherapy \cite{BHARAT2014640, Damato:2014aa, invivo-dosimetry}. On the other hand, the behaviour of PSD is well known, but the presence of a sensor next to it might alter its usual response \cite{scintillation-dosimetry}. 

This study presents an approach where an energy independent dosimeter, a PSD, is coupled to an EM sensor (having intrinsic sub-mm positional accuracy) for real-time dosimeter tracking. A quantification of the added sensor's influence on the signal was made with four different sensors (and cabling). For the most promising one, the sensor's intrinsic error and reproductibility, as well as anglular and positional accuracies were studied. 

\section {Materials and Methods}%=======================
\subsection{Sensor models}

\begin{figure} [H]%working if you put H
\centering
     \includegraphics[width=9cm]{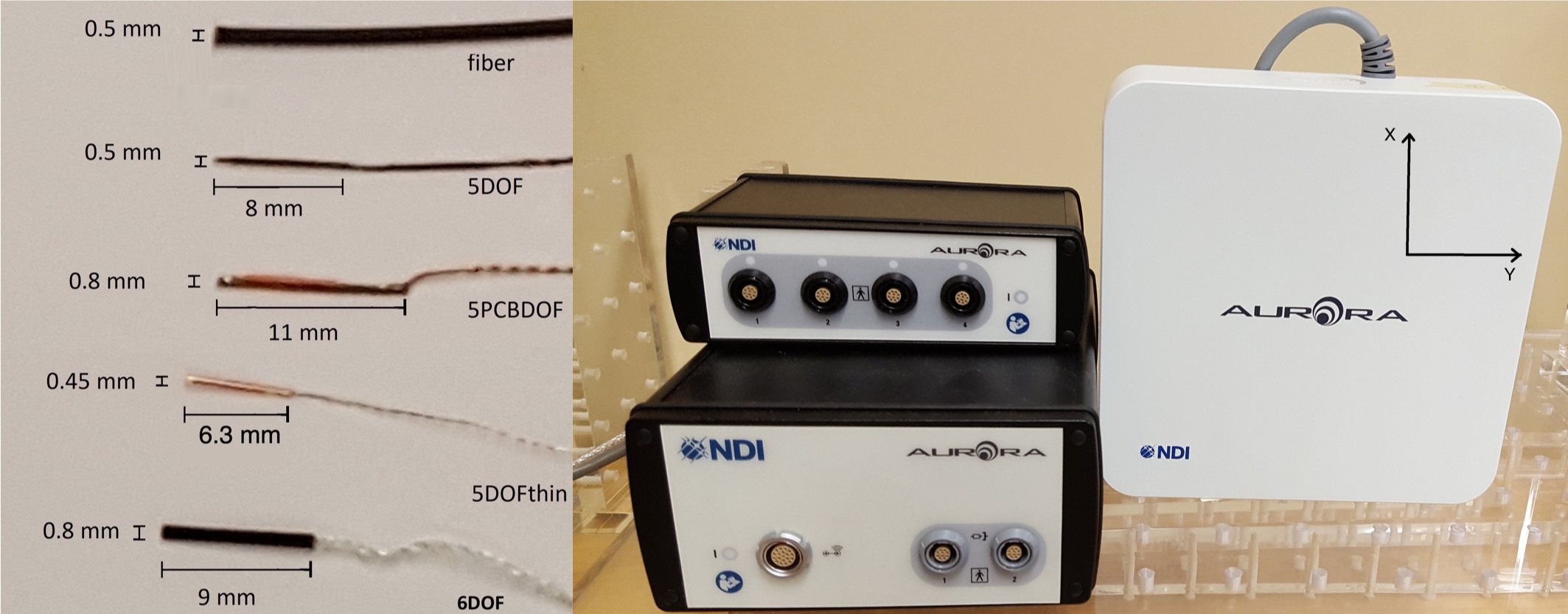}

\caption{Optical fiber, four EM sensors with their collecting cables (left panel) and Aurora V3 electromagnetic tracking system (right panel). These sensors were added to our scintillating fiber dosimeter. Axes are indicated on the field generator, with the Z axis coming out of the figure.
}\label{sensors}
\end{figure}

Each sensor used are presented in Table \ref{sensors-characteristic}. They are also showed on the left panel of Fig. \ref{sensors}. 
\begin{table}[H]
\caption{Characteristic of four EM sensors used with Aurora V3 electromagnetic tracking system.}\label{sensors-characteristic}
\centering
\begin{tabular}{ccccccc}
Sensor&Part Number& Diameter (mm)&Length (mm)\\
\hline
5DOF&610099&0.5&8\\
5PCBDOF&610090&0.8&11\\
5DOFthin&610157&0.45&6\\
6DOF&610059&0.8&9\\

\hline
\end{tabular}
\end{table}
All sensor's cable are twisted-pair copper cables. The 5DOF and the 6DOF sensor are covered with a polyimide  tube to hold and protect the thin sensor and its cable. Each sensor is physically presented in the left panel of Fig. 1.

\subsection{Signal Attenuation}
A single PSD assembly composed of a 0.5 mm diameter and 3 mm long BCF-60 (Saint-Gobain Crystal, Courbevoie, France) and 8 m clear collecting fiber was used with each EM sensor. The EM sensors were read at a rate of 40 sample per second (40Hz) with the Aurora V3 system (NDI, Ontario, Canada) as shown in Fig. 1.  The PSD signal was acquired using a Hamamatsu photomultiplier tube (series H10722) under irradiation at 120 kVp with an orthovoltage unit (Xshrahl 200). This ensures there is no Cerenkov produced and it maximizes the effect of the sensors due to photoelectric effect.

In order to keep the system at the smallest possible diameter, the configuration needs to have the sensor first followed by the scintillator and the light collecting fiber. This arrangement is the envisioned final assembly of a PSD+EMT \textit{in vivo} system having an overall diameter of 0.5 mm and thus one that could be inserted in any needle, catheter or applicator currently used in high-dose rate brachytherapy

To evaluate the influence of distance between the sensor and the PSD on the scintillation, 5 different spacings (0 cm, 1 cm, 2 cm, 3 cm and 5 cm) were tested. In order to keep the expected dose at the scintillator constant, the sensor was moved away from the scintillator while keeping the source-PSD distance fixed. In this test, the cable was placed behind the PSD to avoid any additional scintillating signal attenuation. All the measurements were made at 0$\degree$ (parallel)  and 90$\degree$ (perpendicular). 

Secondly, the shadowing effect from the sensor on the PSD signal was characterized for an optimal sensor-to-PSD distance (found from the measurements above) by taking measurements from 0$\degree$  to 90$\degree$, at increment of 10$\degree$. 

Finally, the effect of the sensor cable on the PSD signal was characterized by having each sensor positioned at 2 cm in front of the PSD and its cable running either in front or in the back, relative to the radiation beam, of the PSD assembly.

\subsection{Sensor Performance Characterization}
Based on the above results, the best performing sensor has been further characterized by its average jitter error and reproducibility for a fixed position i.e. the sensor intrinsic accuracy. With 9 measurements of 10 s, the standard deviation of the average measurement values was taken as the sensor's reproductibility. The jitter error was the average standard deviation of each individual measurement of 10 s (400 samples/measurement). 

Using an acrylic plate and a catheter template, the distance between various positions within the field generator effective volume was measured \cite{hummel2005}. Measurements were taken at every 10 mm, 50 mm and 110 mm on the acrylic plate. Every distances were measured with a dial caliper ($\pm$ 0.1 mm ). A similar study was made for the angles using a machined acrylic support with 10 predefined needle orientation holes ( 2$\degree$, 5$\degree$, 10$\degree$, 15$\degree$, 20$\degree$, 30$\degree$, 45$\degree$, 60$\degree$ and 90$\degree$) inside of which the sensor was placed. The support is the same than the one used in Boutaleb et all.\cite{Boutaleb:2014}. The uncertainty on the needle hole angle is $\pm 1 \degree$. Intermediate angles were obtained by raising one end of the acrylic support at known heights.

\section{Results and Discussion}%========================
\subsection{Signal Attenuation}
\begin{figure} [H]%working if you put H
\centering
     \includegraphics[width=12cm]{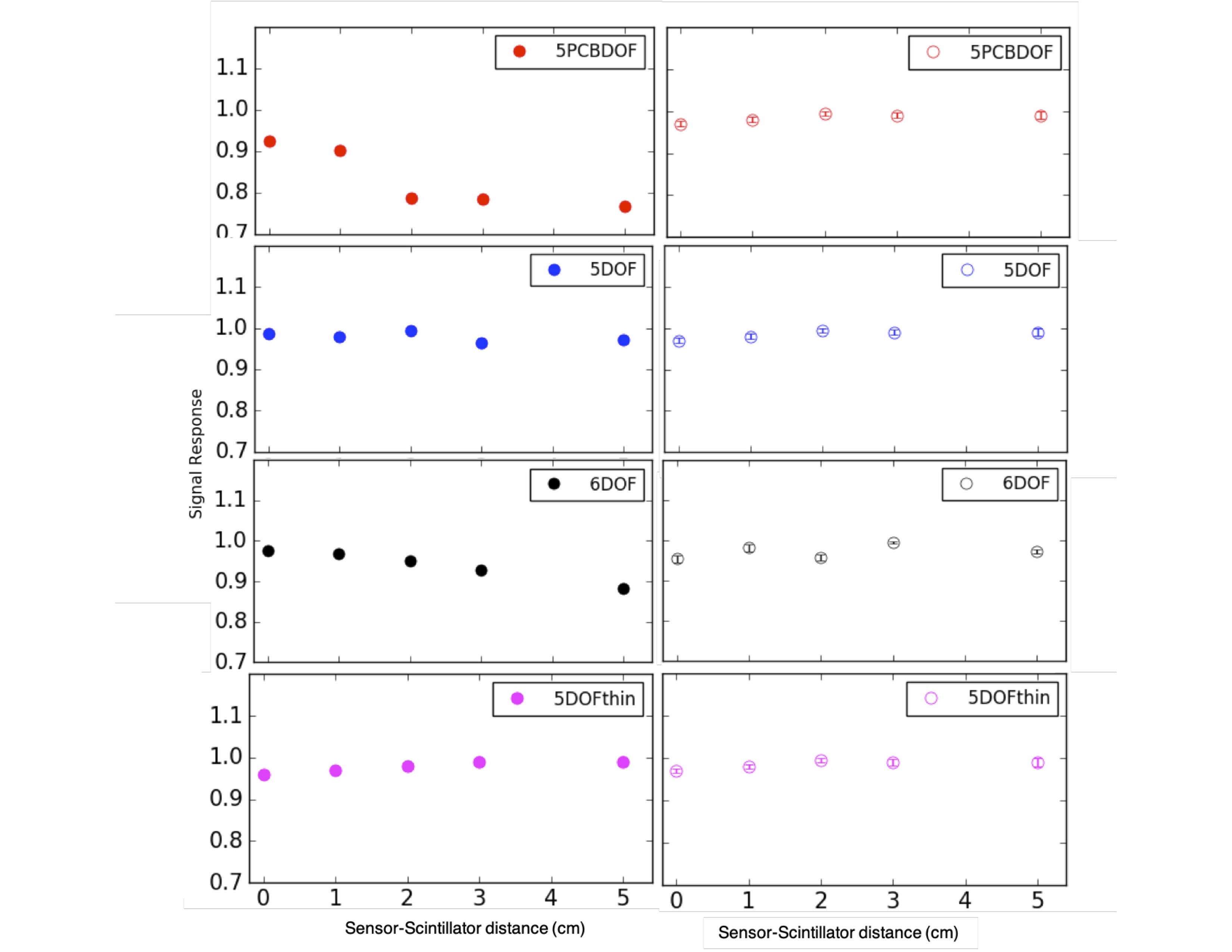}
\caption{Signal dependence on the spacing at 0$\degree$ (left panel) and 90$\degree$ (right panel) for different sensors. Data normalized by number of monitor unit.}\label{distance_effect}
\end{figure}

 Figure \ref{distance_effect} shows an attenuation of the signal when the beam is directly on the dosimeter's  axis (0$\degree$). This effect increases as the distance sensor-PSD increases for 5PCBDOF sensor, thus having a sensor closer to the radiation source and increasing the geometry shadowing of the radiation beam. This sensor is also the thicker one. Note that the 6DOF, the second thicker sensor also exhibits a change in signal at 0$\degree$ as a function of distance. 
 
 At 90$\degree$, no shadowing effect was seen, as expected, and potential scattering effect from the sensor appeared non-existent. Still to be safe, a distance between the sensor and the PSD of 2 cm was selected for all other measurements presented in this manuscript. 
The 5DOF and the 5DOFthin shows little variation with distances and no significant differences between 0$\degree$ and 90$\degree$.

\begin{figure} [H]%working if you put H
\centering
     \includegraphics[width=9cm]{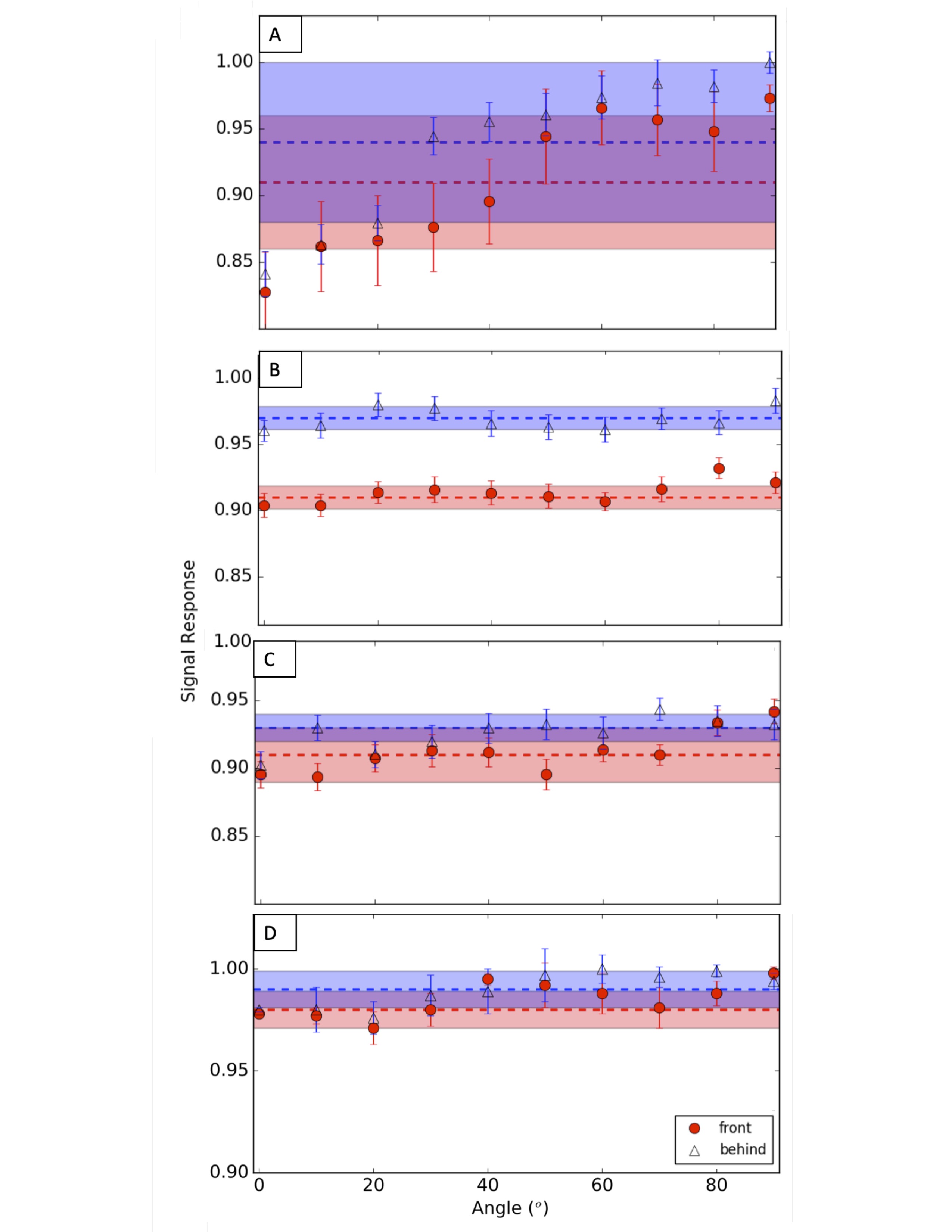}
\caption{Dependence of the scintillator signal on the beam incident angle for 4 different sensors with the cable in front (full dots) and behind (open triangles) the PSD. The data are normalized by the number of monitor unit. The dashed line are the signal average and the colored area its standard deviation. The various sensors are: A) 5PCBDOF B) 5DOF C) 6DOF and D) 5DOFthin}\label{angle_effect}
\end{figure}
In figure \ref{angle_effect}a, the scintillation light intensity increases by almost 20 $\%$ with increasing beam angle, showing a strong angular dependence of the attenuation by the 5PCBDOF sensor for photon coming directly at the tip of the dosimeter (Fig. \ref{angle_effect}), which is in good agreement with the previous results for a distance of 2 cm. Furthermore, the shadowing effect explains the large standard deviation of the signal. Comparing the 2 sets of data for cable positioning (front and behind) gives up to 6 $\%$ difference between them. 
  
For the 5DOF and 6DOF sensors, the small standard deviation highlights that they have essentially no angular dependence. However, the former shows a clear separation between the data based on sensor's cable position, which is not the case for the latter. The 6DOF sensor leads to less PSD signal, thus stronger attenuation from the cable than the 5DOF.
 
The final sensor, the 5DOFthin, shows the best results in terms of scintillation light attenuation with angle as well as minimal cable attenuation. This one is the latest made available to researchers by NDI and because of its size and angular/cable positioning tests, it was adopted for the remainder of this study. 

\subsection{Sensor performance characterization}
The 5DOFthin sensor was characterized by its average jitter error and reproductibility, which were $\pm$ 0.06 mm and $\pm$ 0.008 mm respectively. The errors at different areas in the detection volume are similar to those that Boutaleb et all. presented \cite{Boutaleb:2014}.

\begin{figure} [H]%working if you put H
\centering
     \includegraphics[width=14cm]{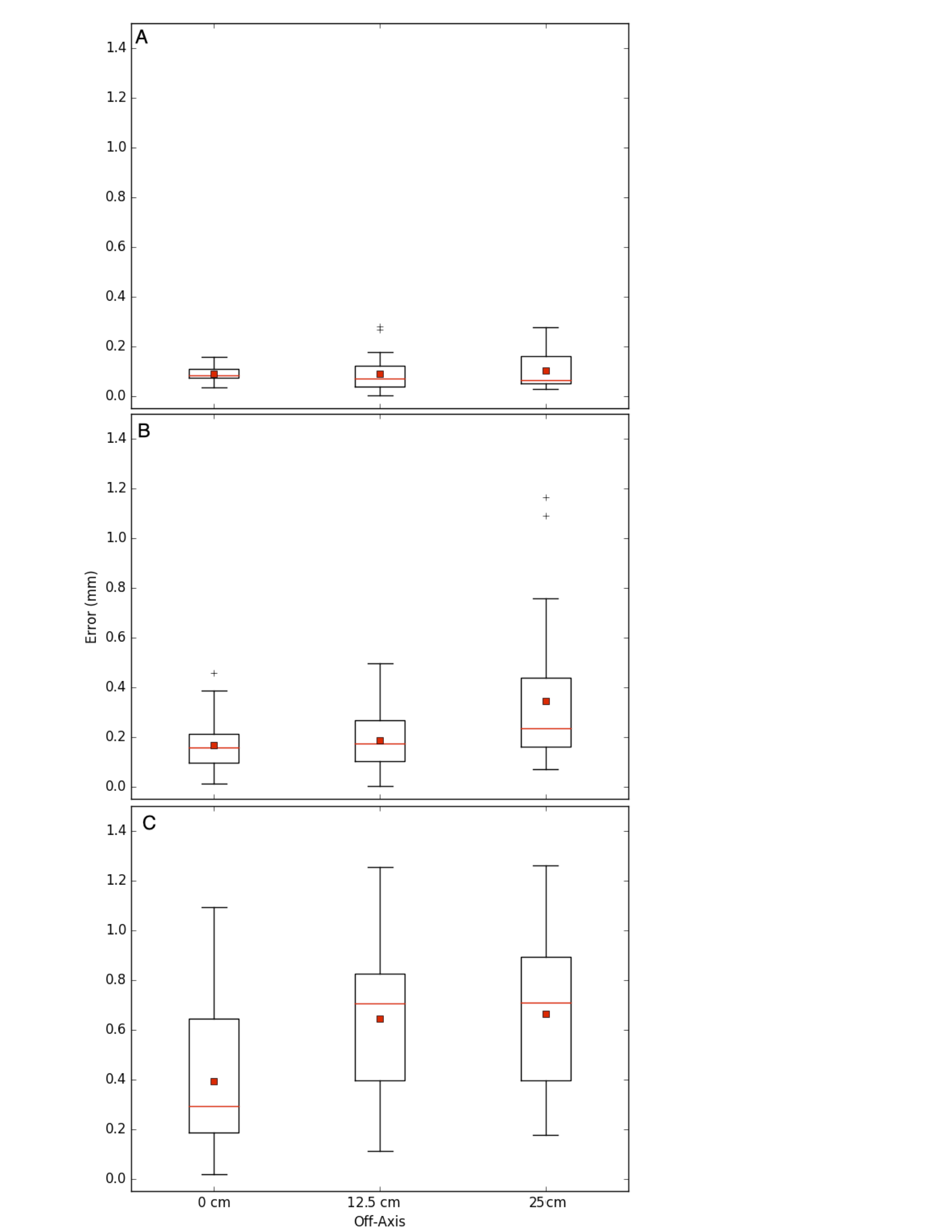}

\caption{Distribution of positional errors as a function of the distance from the generator surface for different off-axis distances from the generator center, namely A) 50 mm B) 250 mm, C) 500 mm and different off-axis distances. The means are presented as squares, while the lines represent the median values. }\label{relative-distance}
\end{figure}

As the sensor moves away from the generator, the relative error increased (see Fig.\ref{relative-distance}). At 5 cm from the generator, the maximum error was less than 0.2 mm  at a position centered in the XY plane. Although, at 25 cm off-axis, the average maximum error increased to 0.3 mm. 

Figure \ref{relative-distance} shows that for the optimal usage distance as defined by Boutaleb et al. \cite{Boutaleb:2014} corresponding to Figure \ref{relative-distance} a and b, all average errors are well under 0.2 mm up to 12.5 cm off-axis. The much larger errors occurring at the largest distances (at the edges) from the field generator. For the optimal central region, using the known dose gradient from a Ir-192 source, an error of 0.2 mm translates to a dosimetric uncertainty of less than 5 \% at 10 mm from the source. These measurements indicates that the small differences between the computed and expected relative distances are precise enough to use in a \textit{in vivo} scintillation dosimeter.

\begin{figure} [H]%working if you put H
\centering
     \includegraphics[width=12cm]{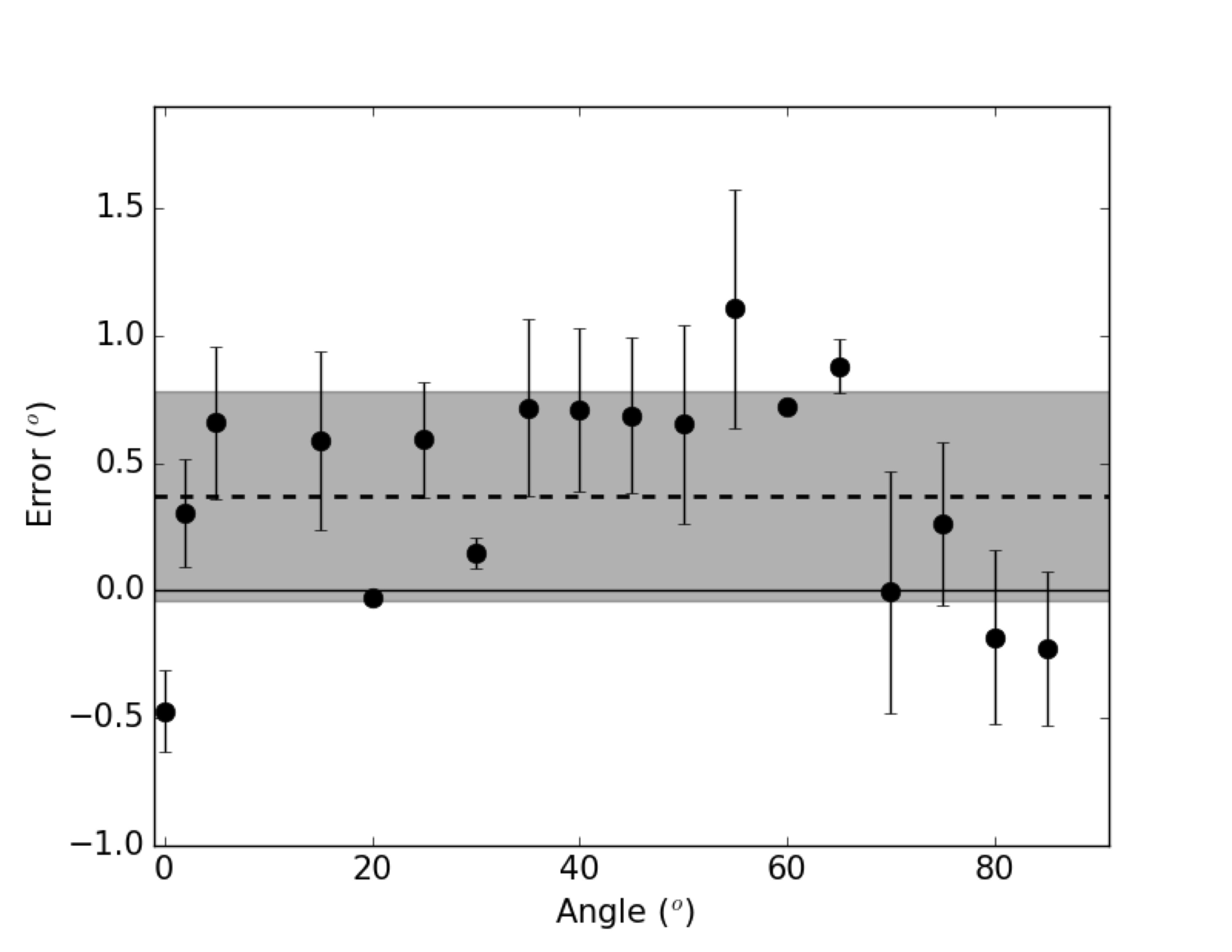}

\caption{Difference between measured and expected values of the 5DOFthin sensor rotation angles. The dotted-line represent the average error and the shaded part, its standard deviation.}\label{relative-angle}
\end{figure}

According to Figure \ref{relative-angle}, the maximum difference between measured and expected values for sensor angulation is 1.1$\degree$. Most of the angles measured are larger than the ones expected. This is due to the insertion hole being slightly larger than the sensor, with the cable pulling the sensor up in the hole, explaining the small but positive shifts.

The error bars are at most of 0.5 $\degree$, which occurs at the 70$\degree$ rotation which is a little higher than the 0.4$\degree$ given by the manufacturer. This can be explained by the $\pm 1\degree$ uncertainty of every needle hole. 

There is another 4D non-water equivalent \textit{in vivo} dosimetry system, RADPOS, which is using an EM tracking system, Ascension (now a NDI company), based on direct continuous current\cite{Cherpak:2009aa}. The overall Aurora AC-based EM tracking system's performance present in this study appears to be better than the DC-system, in line with the results reported in the review by Franz \textit{et al.}. The electromagnetic sensor used in that previous work was a larger 8 mm length cylinder with a 1.3 mm diameter\cite{Cherpak:2009aa}. A sensor-dosimeter distance of 8 mm was used to avoid radiation disturbance, on a similar fashion as the 2 cm sensor-PSD distance used in this work. Cherpak \textit{et al.} tested the accuracy of the RADPOS system and found that between 150 mm to 470 mm distance from the generator center leads to 1.10$ \pm$ 0.07 mm \cite{Cherpak:2009aa,Franz:2014aa}. It is important to underline once more that due to the dose gradient in Ir-192 brachytherapy, a 1 mm uncertainty lead to a dosimetric uncertainty of 55 \% at 5 mm from the source and reach 5\% only at 50 mm. Thus, the positional accuracy reported here by moving to the current Aurora system and novel sensor represents a significant improvement.

 \section{Conclusion}%==============================
This work demonstrates that it is possible to integrate an EM tracking sensor to an energy independent plastic scintillation dosimeter with minimal impact to the collected signal. The sensor chosen, 5DOFthin, has the needed positional accuracy, angular accuracy, and reproducibility for real-time dosimeter tracking. It opens up the possibility to increase the accuracy of \textit{in vivo} dosimetry in Ir-192 high-dose rate brachytherapy by greatly decreasing the uncertainty on the dosimeter position from the overall uncertainty budget.

 %===============================
\section*{Acknowledgments}
This work was supported by the National Sciences and Engineering Research Council of Canada (NSERC) via the NSERC-Elekta Industrial Research Chair. Daline Tho acknowledges support from the Medical Physics Training Network CREATE NSERC grant \# 432290.
 
\bibliographystyle{plain}
\bibliography{matrise}

\end{document}